\documentclass[aps,prb,twocolumn,superscriptaddress]{revtex4-2}
\usepackage{amsmath,amssymb,graphicx,bm}
\usepackage[colorlinks=true,linkcolor=blue,citecolor=blue,urlcolor=blue]{hyperref}
\usepackage{tikz-feynman}
\usepackage{subcaption}
\tikzfeynmanset{compat=1.1.0}

\usepackage{ulem}

\begin{document}

\title{Spin-stripes in the Hubbard model: a combined DMFT and Bethe-Salpeter analysis}

\author{Ruslan Mushkaev}
\affiliation{Department of Physics, University of Fribourg, 1700, Fribourg, Switzerland}

\author{Francesco Petocchi}
\affiliation{Department of Physics, University of Fribourg, 1700, Fribourg, Switzerland}

\author{Shintaro Hoshino}
\affiliation{Department of Physics, Chiba University, Chiba 263-8522, Japan}

\author{Philipp Werner}
\affiliation{Department of Physics, University of Fribourg, 1700, Fribourg, Switzerland}

\date{\today}

\begin{abstract}
    The Hubbard model is known to accommodate various electronic orders, including stripes, which are important for understanding the physics of cuprates. 
    We study spin-stripe order in the square lattice Hubbard model as a function of doping and temperature, by solving the Bethe-Salpeter equation with the local vertex from dynamical mean field theory (DMFT), both inside and outside the antiferromagnetic phase. We find broad regions of horizontal/vertical spin stripes at low temperatures for the model with and without next-nearest neighbor hopping. Their wavelength depends on hole doping in a nonlinear fashion, and is highly sensitive to the ratio of nearest and next-nearest neighbor hoppings.
\end{abstract}

\maketitle

\section{Introduction}

The Hubbard model \cite{hubbard} on the square lattice is the prototypical model for describing the electronic properties of cuprates and related strongly correlated materials \cite{Arovas2022}. The Hamiltonian of the one-band Hubbard model with 
nearest-neighbor hopping $t$, next-nearest neighbor hopping $t'$ and  on-site Coulomb repulsion $U$ reads
\begin{align}
    H = \!-t \sum_{\langle ij \rangle,\sigma} c^\dag_{i\sigma}c^{}_{j\sigma} \!-t' \sum_{\langle \langle ij \rangle \rangle,\sigma} c^\dag_{i\sigma}c^{}_{j\sigma} \!+ U \sum_i n_{i\uparrow} n_{i\downarrow}.
\end{align}
Here, $c^\dag_{i\sigma}$ ($c^{}_{i\sigma}$) creates (annihilates) an electron with spin $\sigma$ at site $i$ and $n_{i\sigma}=c^\dag_{i\sigma}c^{}_{i\sigma}$ measures the spin density. Despite its simplicity, the interplay between the kinetic and interaction energy scales in this model, as well as the effects of filling and temperature, result in a remarkable richness of phenomena, including a Mott transition, Fermi arcs, non-Fermi liquid properties, superconductivity and spin/charge orders \cite{dmft_georges,Lichtenstein2000,fermi_arcs_rubtsov,Haule2007,Kancharla2008,Werner2016,vqmc_ido,stripe_science,stripe_idmft,stripe_prr}. The long-standing question of how to characterize and understand the metallic phase obtained upon doping the Mott insulator away from half-filling \cite{mott_doping} is believed to be crucial for understanding the high-$T_c$ superconductivity in cuprates. For small values of hole doping $\delta \lesssim 0.2$, horizontal/vertical modulations in the charge and spin density of the square lattice -- so called \textit{stripe} states -- have been experimentally observed \cite{tranquada_nature}. Their appearance had been predicted at the end of the 1980s on the basis of Hartree-Fock calculations \cite{stripe_rice,stripe_zaanen}, and more recently confirmed with more advanced many-body techniques such as variational Monte Carlo (VMC) \cite{vqmc_ido}, auxiliary-field quantum Monte Carlo (AFQMC), the density matrix renormalization group (DMRG), density matrix embedding theory (DMET), and infinite projected entangled pair states (iPEPS)  \cite{stripe_science,stripe_prr}, as well as inhomogeneous dynamical mean field theory (iDMFT) \cite{pavarini_stripes,stripe_idmft}. There are, however, inconsistencies between the theoretical predictions and experimental results. For instance, the experimentally observed stripes are half-filled (1/2 hole per unit cell of the charge order), which contradicts the numerically favored filled stripes (one hole per unit cell of the charge order) 
\cite{stripe_science} or 2/3-filled stripes \cite{dmrg_jiang}. This discrepancy has been attributed to the very small energy scales involved in stripe energetics, and also to the inadequacy of the simple Hubbard model treatment, which does not include longer range hoppings and long-range Coulomb repulsions. Furthermore, at present, the interplay of stripes with superconductivity is still unclear. Since stripes break translational and C$_4$ symmetries, they prevent a uniform superconducting phase due to weak inter-chain Josephson coupling between the 1D-like strips \cite{vqmc_ido}, although superconductivity can still develop within the stripe state, if the condensate forms a modulated pair-density wave \cite{tranquada_review}.

In this paper, we focus on mapping out the temperature-doping phase diagram for spin-stripe order for the single-band Hubbard model with and without next-nearest neighbor hopping on a square lattice. By employing the Bethe-Salpeter (BSE) formalism within the framework of dynamical mean-field theory (DMFT) \cite{dmft_georges}, we can compute momentum dependent spin susceptibilities by measuring the local two-particle irreducible particle-hole vertex. The advantage of this method is that it is not prone to finite-size effects, and does not require the biasing of the solution by pinning fields, which restrict the solutions to commensurate ordering vectors. Also, our temperature-dependent data, obtained with a continuous-time QMC solver \cite{werner_ct_hyb}, are beyond the reach of ground-state numerical methods. By analyzing the static spin susceptibility on a high-symmetry path in the Brillouin zone, we study the competition between horizontal [vertical] and diagonal spin stripes, characterized by susceptibility peaks at momenta $\mathbf{q}_\text{stripe}=(\pi,\pi-2 \pi \eta_{\mathrm{hor}})$ [$(\pi-2 \pi \eta_{\mathrm{hor}},\pi)$] and $(\pi-2 \pi \eta_{\mathrm{diag}},\pi-2 \pi \eta_{\mathrm{diag}})$, respectively, and map out the phase boundary corresponding to the leading divergence. 

This paper is structured as follows: section II summarizes the DMFT-BSE formalism in the relevant particle-hole (PH) channel, section III presents the main results in the form of a temperature-doping phase diagram and susceptibility plots, and section IV summarizes the most important findings. A detailed description of the susceptibility measurements can be found in the appendices.

\section{Method}

\subsection{DMFT-BSE formalism}

DMFT is a Green's function based method, where correlations are encoded in a local, but frequency-dependent self-energy \cite{dmft_georges}. The lattice system is mapped onto an interacting impurity, embedded into a self-consistently determined bath. DMFT enables one to measure the local two-particle irreducible vertex, which can be inserted into the particle-hole Bethe-Salpeter equation to obtain the particle-hole susceptibility. The DMFT-BSE equations for the two-particle Green's functions of the lattice and impurity system read:
\begin{equation}
    \begin{aligned}
        &\underline{\chi}^{\mathrm{PH}}(Q) =  \underline \chi^{0,\mathrm{PH}}(Q) + \frac{1}{\beta^2} \underline \chi^{0,\mathrm{PH}}(Q) \underline{\tilde{\Gamma}}^{\mathrm{PH}}(i \omega_n) \underline \chi^{\mathrm{PH}}(Q),
    \end{aligned}
    \label{bse_lat}
\end{equation}
\begin{equation}
    \begin{aligned}
        &\underline{\tilde{\chi}}^{\mathrm{PH}}(i \omega_n) =  \underline{\tilde{\chi}}^{0,\mathrm{PH}}(i \omega_n) + \frac{1}{\beta^2} \underline{\tilde{\chi}}^{0,\mathrm{PH}}(i \omega_n) \underline{\tilde{\Gamma}}^{\mathrm{PH}}(i \omega_n)\underline{\tilde{\chi}}^{\mathrm{PH}}(i \omega_n),
    \end{aligned}
    \label{bse_loc}
\end{equation}
where $Q = (i\omega_n,\mathbf{q})$ is a combined index for the Matsubara frequencies and momenta, and $\omega_n=\frac{2 n \pi}{\beta}$ (with $\beta$ the inverse temperature of the system) are bosonic Matsubara frequencies. The irreducible vertex $\underline{\tilde{\Gamma}}^{\mathrm{PH}}$ is obtained from the local (impurity) two-particle Green's function by solving the local BSE in Eq.~\eqref{bse_loc}. This local impurity vertex is then used in the lattice BSE, Eq.~\eqref{bse_lat}, to define the particle-hole susceptibility $\underline{\chi}^{\mathrm{PH}}(Q)$. The underline denotes a matrix in fermionic frequencies $\nu_n = \frac{(2n+1)\pi}{\beta}$ and spin-sublattice indices, with the following indexing:
\begin{align}
    A^{\mathrm{PH}}_{abcd}(i\nu_n,i\nu_n') \rightarrow \underline A^{\mathrm{PH}}_{(ab,i\nu_n)(dc,i\nu_n')},
\end{align}
where the two-particle Green's function $A_{abcd}$ on the left-hand side will be specified below.

We define the non-interacting particle-hole bubbles of the impurity and lattice, respectively, as
\begin{align}
    &\underline{\tilde{\chi}}^{0,\mathrm{PH}}(i \omega_n)_{(\bar{a}b,i \nu_n),(d\bar{c},i \nu_n')} \nonumber\\ 
    &= - \tilde{G}_{d \bar{a}}(i\nu_n) \tilde{G}_{b \bar{c}}(i\nu_n-i \omega_n) \delta_{i\nu_n,i\nu_n'} , \\ 
    &\underline{\chi}^{0,\mathrm{PH}}(Q)_{(\bar{a}b,i \nu_n),(d \bar{c},i \nu_n')} \nonumber\\ 
    &= - \frac{1}{N_k} \sum_{\mathbf{k}} G_{d \bar{a}}(\mathbf{k},i\nu_n) G_{b \bar{c}}(\mathbf{k}-\mathbf{q},i\nu_n - i \omega_n) \delta_{i\nu_n,i\nu_n'} ,
\end{align}
where barred (unbarred) indices refer to the creation (annihilation) operators. The connected particle-hole two-particle Green's function of the impurity is given by
    \begin{align}
        &\tilde{\chi}^{\mathrm{PH}}(i \omega_n)_{(\bar{a}b,i \nu_n),(d \bar{c},i \nu_n')} \nonumber\\ 
        &= \tilde{G}^{\mathrm{PH}}_{\bar{a}b\bar{c}d}(i\omega_n,i\nu_n,i\nu_n') - \delta_{i \omega_n,0} \tilde{G}_{b \bar{a}}(i\nu_n) \tilde{G}_{d \bar{c}}(i\nu_n'),
    \end{align}
with the local two-particle Green's function defined as 
    \begin{align}
        &\tilde{G}^{\mathrm{PH}}_{\bar{a}b\bar{c}d}(i\omega_n,i\nu_n,i\nu_n') = \frac{1}{\beta^2} \iiiint_0^\beta d \tau_1 d \tau_2 d \tau_3 d \tau_4 \nonumber\\ 
        & \times e^{-i \nu_n(\tau_1-\tau_2)} e^{-i \omega_n(\tau_2-\tau_3)} e^{-i \nu_n'(\tau_3-\tau_4)} \nonumber\\ 
        & \times \langle \mathcal{T}_\tau \hat{c}^{\dag}_{\bar{a}}(\tau_1) \hat{c}_b(\tau_2) \hat{c}^{\dag}_{\bar{c}}(\tau_3) \hat{c}_d(\tau_4) \rangle.
    \end{align}

Equations \eqref{bse_lat} and \eqref{bse_loc} can be combined into the expression
\begin{align}
    &\underline{\chi}^{\mathrm{PH}}(Q) = \nonumber\\ 
    &\hspace{5mm}\Bigg[ \underline{\chi}^{(0,\mathrm{PH})-1}(Q) - \underline{\tilde{\chi}}^{(0,\mathrm{PH})-1}(i \omega_n) + \underline{\tilde{\chi}}^{(\mathrm{PH})-1}(i \omega_n) \Bigg]^{-1} , 
\end{align}
which yields the physical spin susceptibility $\chi_{OO}(Q)$ for a Hermitian spin observable $O$ through the relation
\begin{align}
    \chi_{OO}(Q) =  \frac{1}{\beta} \sum_{i\nu_n,i\nu_n'}  \sum_{\bar{a} b \bar{c} d} O_{\bar{a}b} O_{\bar{c}d} \chi^{\mathrm{PH}}_{\bar{a}b\bar{c}d}(Q,i\nu_n,i\nu_n'). 
\end{align}
We will use $O_{ab}=O_{aa} \delta_{ab}$ and $O_{aa} = O_{(\lambda,\sigma)(\lambda,\sigma)} = (-1)^{(1-\sigma)}$ for $zz$-spin correlations ($\lambda$ and $\sigma=1,2$ are sublattice and spin indices, respectively).

If the system is in the antiferromagnetic (AFM) phase, the form of the BSE has to be adapted to this long-range order. In practice, we then perform the calculations with a doubled unit cell and explicitly break the SU(2) symmetry. Details of the formulation of the DMFT-BSE equations outside and inside the AFM-ordered phase are provided in Appendices ~\ref{bse_para_app} and ~\ref{bse_afm_app}, respectively. Since our DMFT simulations employ diagonal hybridization functions, which restricts the measurable components of the two-particle Green's function in the AFM phase, this means that the lattice BSE has to be solved in a correspondingly restricted spin-sublattice space.  

\subsection{Frequency convergence}

The BSE in Eqs.~(\ref{bse_lat}) and (\ref{bse_loc}) are solved in a finite frequency box with $|\nu_n|,|\nu_n'| \leq \nu_\mathrm{max}$. Upon performing the Matsubara sum, the resulting spin susceptibility scales as $\sim \frac{1}{\nu_\mathrm{max}}$. In practice, for high temperatures, $\beta=1/T \leq 8$, we use a fixed frequency box of 16 positive fermionic frequencies, and for lower temperatures $8 < \beta \leq 24$, we use 32 positive fermionic frequencies and extrapolate to the infinite frequency limit. Although lower temperatures are accessible in DMFT, they require an increasingly large frequency box, which can create problems with high-frequency noise in the inversion of the two-particle quantities \cite{schafer_mott_hubbard}. In practice, we start with a small frequency box and gradually increase the number of frequencies until the quantities of interest are sufficiently converged.

\section{Results and Discussion}

We use the DMFT-BSE formalism described above to investigate the spin ordering patterns in the temperature-doping phase diagram. We set the nearest-neighbor hopping $t=1$ as the energy unit in this paper, and the lattice constant to $a=1$. The hole doping is defined by $\delta = 1-n$ with $n$ the average electron number per site.

\subsection{Temperature-doping phase diagram at $U=8$}

\begin{figure}[t]
    \centering
    \includegraphics[width=\linewidth]{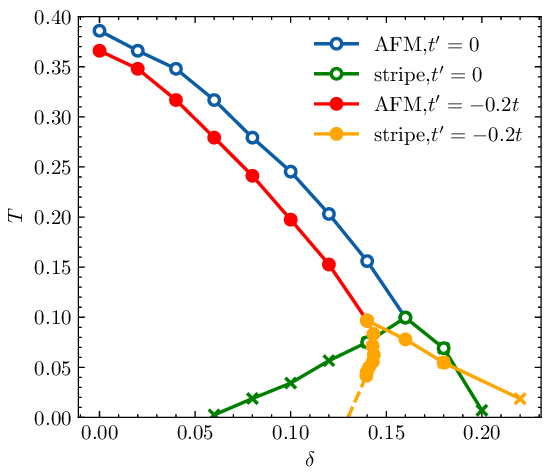}
    \caption{Spin-stripe and AFM order in the Hubbard model with (red, yellow) and without (blue, green) next-nearest neighbor hopping $t'$ and $U=8$. Circles mark diverging susceptibilities, while crosses are phase transition points obtained by Curie-Weiss extrapolation.
    The dashed yellow line is an extrapolation of the phase boundary.
    }
    \label{phase_diagram}
\end{figure}

The phase diagram at $U=8$ is depicted in Fig.~\ref{phase_diagram}, where all the transition lines have been obtained by detecting a corresponding divergence of the spin-susceptibility at an appropriate momentum point. The half-filled $(\delta=0)$ Hubbard model without next-nearest neighbor hopping ($t'=0$) shows AFM order below a temperature $T\approx 0.4$. As the hole doping $\delta$ increases, the N\'eel temperature decreases, but up to $\delta\approx 0.15$, the temperature driven transition is into a conventional AFM phase. Our AFM transition line is consistent with the DMFT results of Ref.~\cite{dual_fermion_otsuki}.

At doping $\delta=0.16$, the dominant AFM magnetic susceptibility peak at $\mathbf{q}=(\pi,\pi)$ undergoes a splitting, and in the doping range $0.16\le \delta \lesssim 0.20$, the temperature driven transition is into a stripe-ordered phase (green line in Fig.~\ref{phase_diagram}).  The spin stripe phase also extends into the lower doping region, with a transition temperature which becomes progressively lower as $\delta$ decreases. The green transition line for $\delta<0.16$ has been obtained by performing DMFT-BSE calculations in the ordered phase, where we use a doubled unit cell to break the SU(2) symmetry, and unfold the spin-susceptibility to the Brillouin zone of the original one-unit cell square lattice (see Appendix~\ref{bse_afm_app} for details). Cross symbols indicate $T_c$ values estimated from a Curie-Weiss ($\sim 1/T$) fit of the susceptibility maximum. For $\delta > \delta_{\mathrm{cr}} \simeq 0.20 $, the Curie-Weiss temperature becomes negative, which indicates the disappearance of the spin stripe solution. The obtained stripe phase for the $t'=0$ model is in agreement with the ground-state iDMFT and AFQMC results of Refs.~\cite{stripe_idmft,stripe_prr}, which did not report stripes in the filling range $n < 0.8$ $(\delta > 0.2)$ for $U=8$. 

Introducing a next-nearest neighbor hopping $t'=-0.2$ results in a slight suppression of the AFM phase due to frustrated spin interactions and suppressed Fermi-surface nesting near half-filling. While the spin-stripe phase appears to extend to larger doping values, it is strongly suppressed in the low-doping region, compared to the model without $t'$. Within the temperature range studied, we have not observed divergences in the spin susceptibility for $\delta \leq 0.12$. The boundary appears to fall quite sharply at a certain critical doping when approaching half-filling ($\delta \to 0$), unlike in the $t'=0$ model. The yellow dashed line in Fig.~\ref{phase_diagram} is a naive extrapolation of the phase boundary.

\begin{figure}[t]
    \centering
        \includegraphics[width=\linewidth]{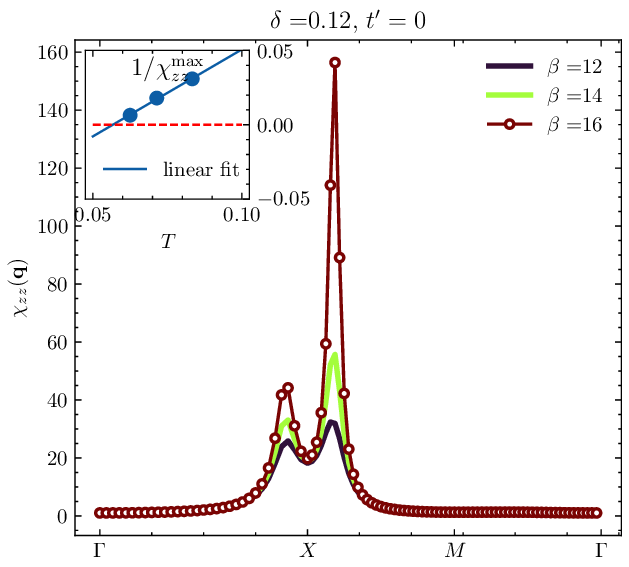}
        \label{chi_zz_t_nn0_delta012}
    \caption{Temperature dependence of the static spin susceptibility in the AFM phase for %two different hole doping values $\delta$.
    hole doping $\delta=0.12$.
    The inset shows the inverse of the susceptibility maximum as a function of temperature, and the Curie-Weiss fit used to extract the ordering temperature. 
    }
    \label{fig:chi_zz_t_nn0}
\end{figure}

\subsection{Spin susceptibility}

We now present the spin susceptibility data which underpin the phase diagram shown in Fig.~\ref{phase_diagram}.
Figure~\ref{fig:chi_zz_t_nn0} depicts the momentum dependence of the spin susceptibility in the AFM phase for the Hubbard model with $t'=0$, for a range of inverse temperatures $\beta=1/T$ but fixed doping. It undergoes a peak splitting around the $X=(\pi,\pi)$ point (in the original Brillouin zone of the square lattice), with the dominant peak along the given path appearing between $X$ and $M=(\pi,0)$. The divergence of this peak represents horizontal spin-stripes, with long wavelength spin modulations in the $k_y$ direction and AFM ordering in the $k_x$ direction. The inset panel depicts the Curie-Weiss fit to the inverse of the susceptibility maxima, and demonstrates that this fit enables a rather precise determination of $T_c$.

\begin{figure}[t]
    \centering
    \includegraphics[width=\linewidth]{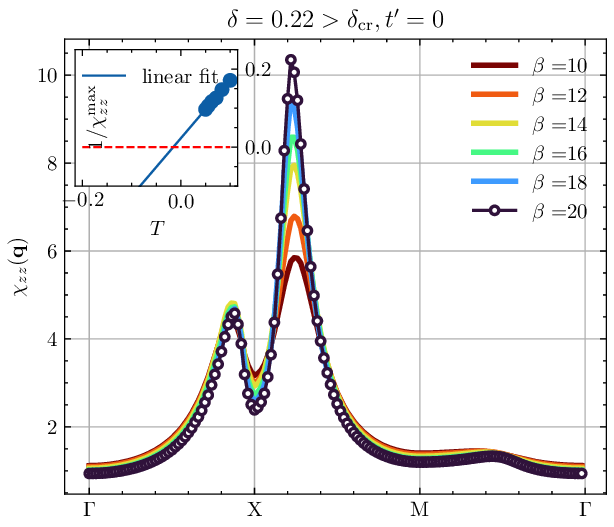}
    \caption{Temperature dependence of the spin susceptibility measured in the paramagnetic phase for $\delta > \delta_{\mathrm{cr}}$ ($t'=0$). The inset demonstrates a negative Curie-Weiss temperature and hence the absence of stripe order.}
    \label{chi_zz_t_nn0_delta022}
\end{figure}

The maximum along the $\Gamma$-$X$ path indicates an enhanced susceptibility for diagonal stripes. However, the horizontal stripe peak always diverges first upon lowering temperature, for all dopings considered. 

In Fig.~\ref{chi_zz_t_nn0_delta022} we show the temperature dependence of the spin susceptibility for $\delta=0.22>\delta_c$, to demonstrate the above-mentioned negative Curie-Weiss temperature, i.e., the absence of stripe order. These susceptibility calculations are performed in the paramagnetic phase.

\begin{figure}[t]
    \centering
    \includegraphics[width=\linewidth]{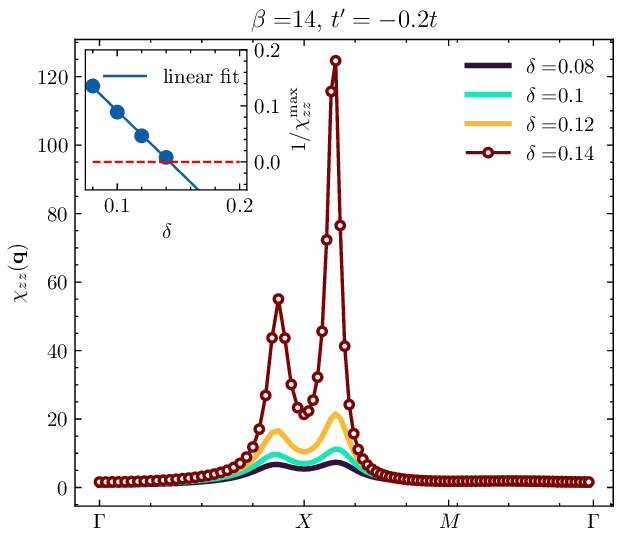}
    \caption{Doping dependence of the static spin susceptibility in the AFM phase for $t'=-0.2$ and $\beta=14$. The inset shows the linear fit to the inverse susceptibility which was used to determine the phase boundary.}
    \label{chi_zz_t_nn02_beta14}
\end{figure}

Figure~\ref{chi_zz_t_nn02_beta14} depicts a scan over a range of dopings for a fixed temperature ($t'=-0.2$). The inverse of the susceptibility maximum displays a linear scaling with $\delta$, as seen in the inset, which can be used to determine the phase transition point. The yellow phase boundary in Fig.~\ref{phase_diagram} between the AFM and stripe phase has been obtained using such doping scans.

\begin{figure}[ht!]
    \centering
    \begin{subfigure}{\linewidth}
        \centering
        \includegraphics[width=\linewidth]{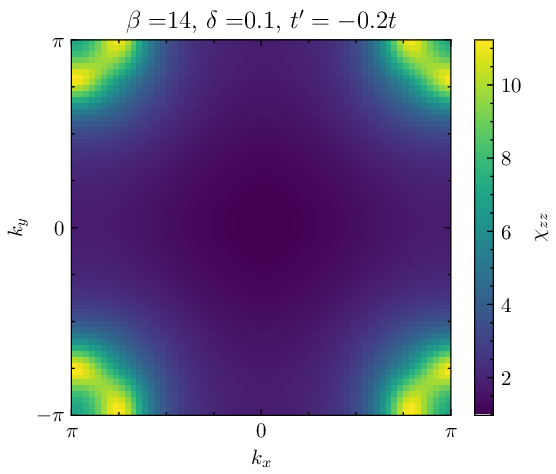}
    \end{subfigure}
    \vspace{0.5cm} 
    \begin{subfigure}{\linewidth}
        \centering
        \includegraphics[width=\linewidth]{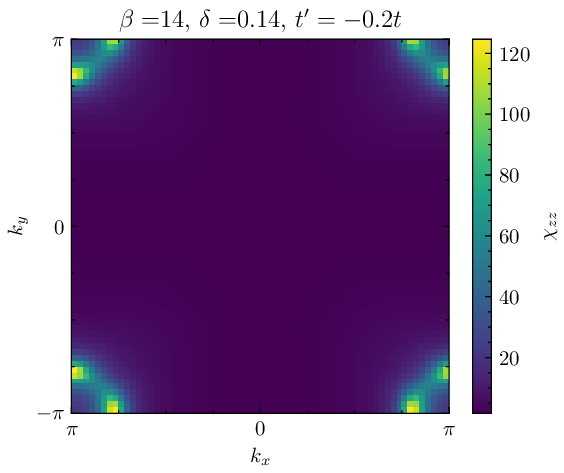}
    \end{subfigure}
    \caption{Full momentum dependence of the static spin susceptibility in the AFM phase for $t'=-0.2$ and two different hole doping values $\delta$. Note the different color bar ranges in the two panels. }
    \label{fig:chi_zz_kplot}
\end{figure}

We also plot the static spin susceptibility ($\omega_n=0$) in the full Brillouin zone of the square lattice in Fig.~\ref{fig:chi_zz_kplot}, both for a small and large doping value. We see that the AFM peak at $(\pi,\pi)$ splits symmetrically along the $(0,\pm \pi),(\pm \pi,0)$ directions. There is hence no properly defined diagonal stripe peak along the $(0,0)$-$(\pi,\pi)$ line, and the lower maximum in Fig.~\ref{chi_zz_t_nn02_beta14} is simply a consequence of the enhanced susceptibilities associated with horizontal and vertical stripes.

\subsection{Comparison with experiment}

It is interesting to investigate the doping dependence of the obtained stripe modulations, and to compare them with experiment. Ref.~\cite{stripes_experiment} presented experimental results on doped cuprates of the La$_{2-x}$Sr$_x$CuO$_4$ family, which can be described by an effective one-band Hubbard model \cite{pavarini_stripes,lorentzana_doping_stripes}. First, let us briefly summarize the predictions from other numerical studies which currently show no consensus. Methods such as AFQMC, DMRG, DMET, iPEPS and iDMFT \cite{stripe_science,stripe_idmft,stripe_prr} predict spin-stripe wavelengths of $\lambda_s = \frac{2}{\delta}$, accompanied by charge-stripe wavelengths of $\lambda_c = \frac{1}{\delta}$, in units of lattice spacing. The factor of 2 difference between the two is a consequence of momentum conservation, which can be seen by applying Landau theory to coupled spin and charge order parameters \cite{landau_theory_stripes}. On the other hand, a $2/3$ CDW and other stripe patterns have been reported in DMRG and VQMC studies \cite{dmrg_jiang,prx_qin,vqmc_ido}. Indeed, due to the small energy scales involved, various stripe orders can be stabilized with different system sizes, making the determination of the true ground state challenging.

Furthermore, these theoretical predictions do not agree well with experiment \cite{stripes_experiment}, which yields $\lambda_s^{\mathrm{exp}} \simeq \frac{1}{\delta}$ for $\delta < \frac{1}{8}$, with a flattening plateau beyond, for Sr-doped cuprates. To our knowledge, the only work which reproduced this experimental behavior is Ref.~\cite{pavarini_stripes}, which employed inhomogeneous (real-space) DMFT, albeit for a rather small number of sites ($\le 10$). These results are however not consistent with the predictions from Ref.~\cite{stripe_idmft}, which implemented iDMFT calculations on larger lattice sizes ranging from 400 to 2000 sites. The lattice size is indeed an important convergence parameter for iDMFT, which can have a profound influence on the wavelengths of the stabilized orders. The latter is not an issue in our DMFT-BSE approach, since we work in momentum space with, in principle, arbitrary momentum resolution. DMFT-BSE calculations can hence probe ordering tendencies with different momentum wavevectors on an equal footing. 

\begin{figure}[t]
    \centering
    \includegraphics[width=\linewidth]{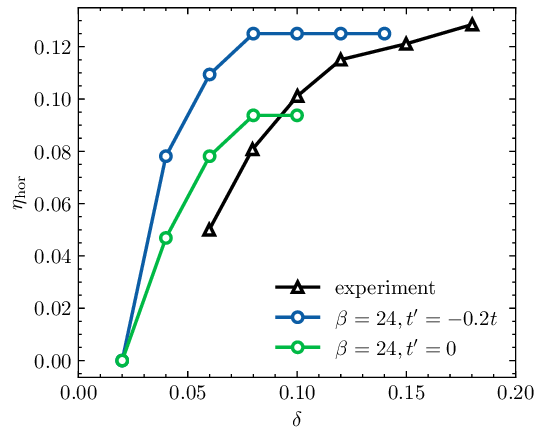}
    \caption{Spin incommensurability parameter for the Hubbard model with and without nearest-neighbor hopping $t'$ compared to experimental data from Ref.~\cite{stripes_experiment}.
    }
    \label{eta_delta_t_t_nn}
\end{figure}

The spin-stripe wavelength can be expressed as a peak-shift $\eta_{\mathrm{hor}}$ in momentum space: $\mathbf{q}_\text{stripe}=(\pi,\pi-2 \eta_{\mathrm{hor}} \pi)$ for the horizontal stripes. This peak shift is plotted against doping in Fig.~\ref{eta_delta_t_t_nn}, for the lowest temperature accessible ($\beta=24$). We first note that $\eta_{\mathrm{hor}}$ depends non-linearly on $\delta$, unlike the results of the large-size iDMFT calculations \cite{stripe_idmft}. We further observe that a next-nearest neighbor hopping $t'=-0.2$, which is a realistic value for the La$_{2-x}$Sr$_x$CuO$_4$ compounds \cite{lorentzana_doping_stripes}, results in $\eta_{\mathrm{hor}}$ values which are consistent with experiment, at least for larger dopings. A lock-in effect, where $\eta_{\mathrm{hor}}$ tends to a constant value is observed at around $\delta \simeq 0.08$, similar to that reported in Ref.~\cite{pavarini_stripes} for the model without $t'$. 

The influence of the ratio of $|t'/t|$ on the stripe filling has already been investigated in Ref.~\cite{lorentzana_stability}, where it was found that this ratio plays a key role by modifying the stripe energetics through constructive and destructive interference of nearest and next-nearest neighbor hopping processes. In particular, it was found that the stripe filling fraction $\nu = \lambda_c \delta$, \cite{lorentzana_doping_stripes} defined as the total doping within one unit cell of the charge order, decreases with increasing $|t'/t|$. Since $\lambda_s = 2 \lambda_c$, the latter implies $\eta_{\mathrm{hor}} = \frac{\delta}{2 \nu}$, i.e. that $\eta_{\mathrm{hor}}$ should increase with decreasing $\nu$. This is indeed the trend observed in Fig.~\ref{eta_delta_t_t_nn}, where $\eta_{\mathrm{hor}}$ is shifted to higher values upon introducing next-nearest neighbor hopping. 

\section{Conclusions}
We investigated spin-stripe order in the square-lattice Hubbard model, by solving the Bethe-Salpeter equation with the DMFT vertex in and out of the AFM phase. We found a horizontal/vertical-stripe solution over a wide range of dopings at low temperatures and this solution becomes the only spin-ordered phase at large doping. Diagonal stripe tendencies were also found, but they are simply a consequence of the enhanced susceptibilities for horizontal and vertical stripes, which are the dominant instabilities for the models with and without next-nearest neighbor hopping. For $t'=0$, the extent of the spin-stripe region in the doping space agrees with the zero temperature iDMFT results of Ref.~\cite{stripe_idmft}. 

Introducing a moderate negative next-nearest neighbor hopping $t'$ significantly alters the stripe filling, and pushes the stripe region to higher doping values in the phase diagram. Choosing a realistic value of $t'$ improves the qualitative agreement of the spin incommensurability parameter with experiment, at least in the optimally doped regime. 

We also briefly comment on the issue of phase separation, which has not been studied here, since our DMFT setup assumes a uniform or AFM state. Phase separation between AFM and stripe orders has been reported in the region $0 < \delta < 1/8$ for $U = 10$ in Ref.~\cite{vqmc_ido}, and for $0 < \delta < 0.15$ for $U=8$ in Ref.~\cite{dual_fermion_otsuki}, adding additional complexity to the temperature-doping phase diagram.

Lastly, although DMFT overestimates the AFM ordering tendency and likely also the spin-stripe transition temperature, the doping ranges for the spin-stripe solutions in both our models overlap with the superconducting regions of the phase diagram for doped cuprates \cite{cuprates_review_armitage}, suggesting a nontrivial interplay between the two phases. Interestingly, superconducting correlations inside the stripe region were recently demonstrated in Ref.~\cite{stripes_sc_science_xu}. 

On a broader level, our study shows that the DMFT-BSE approach is a useful method for tracking electronic ordering instabilities in correlated lattice systems.

\begin{acknowledgments}
The calculations were run on the beo05 cluster at the University of Fribourg. RM and PW acknowledge support by the Swiss National Science Foundation through NCCR Marvel. 
\end{acknowledgments}

\appendix
\begin{widetext}
\section{DMFT-BSE equations in the paramagnetic phase}
\label{bse_para_app}
\subsection{General form}

Our local two-particle vertex is obtained from DMFT using a solver which only treats diagonal hybridizations \cite{werner_ct_hyb}. Hence, we are only able to capture a subset of the local two-particle correlations, with pairwise equal spin-orbital indices. For the particle-hole channel, we have access to the following quantities:
\begin{align}
    \tilde{G}^{\mathrm{PH1}}_{ab}(i\omega_n,i\nu_n,i\nu_n') &:= \tilde{G}^{\mathrm{PH}}_{\bar{a}a\bar{b}b}(i\omega_n,i\nu_n,i\nu_n'),  \\ 
    \tilde{G}^{\mathrm{PH2}}_{ab}(i\omega_n,i\nu_n,i\nu_n') &:= \tilde{G}^{\mathrm{PH}}_{\bar{a}b\bar{b}a}(i\omega_n,i\nu_n,i\nu_n') , \quad a \neq b .
\end{align}
We consider the PH1 channel, which gives us access to the $zz$-spin and charge susceptibilities. The Bethe-Salpeter equations read
\begin{equation}
    \begin{aligned}
        &\underline{\chi}^{\mathrm{PH}}(Q) =  \underline \chi^{0,\mathrm{PH}}(Q) + \frac{1}{\beta^2} \underline \chi^{0,\mathrm{PH}}(Q) \underline{\tilde{\Gamma}}^{\mathrm{PH}}(i \omega_n) \underline \chi^{\mathrm{PH}}(Q) , 
    \end{aligned}
\end{equation}
\begin{equation}
    \begin{aligned}
        &\underline{\tilde{\chi}}^{\mathrm{PH}}(i \omega_n) =  \underline{\tilde{\chi}}^{0,\mathrm{PH}}(i \omega_n) + \frac{1}{\beta^2} \underline{\tilde{\chi}}^{0,\mathrm{PH}}(i \omega_n) \underline{\tilde{\Gamma}}^{\mathrm{PH}}(i \omega_n)\underline{\tilde{\chi}}^{\mathrm{PH}}(i \omega_n) , 
    \end{aligned}
\end{equation}
where 
\begin{align}
    \tilde{\chi}^{0,\mathrm{PH1}}(i \omega_n)_{(a,i \nu_n),(b,i \nu_n')} &= - \tilde{G}_{a}(i\nu_n) \tilde{G}_{a}(i\nu_n-i \omega_n) \delta_{i\nu_n,i\nu_n'} \delta_{ab}, \\ 
    \chi^{0,\mathrm{PH}}(\mathbf{q},i \omega_n)_{(a,i \nu_n),(b,i \nu_n')} &= - \frac{1}{N_k} \sum_{\mathbf{k}} G_{b \bar{a}}(\mathbf{k},i\nu_n) G_{a \bar{b}}(\mathbf{k}-\mathbf{q},i\nu_n - i \omega_n) \delta_{i\nu_n,i\nu_n'}, \\  
    \tilde{\chi}^{\mathrm{PH1}}(i \omega_n)_{(a,i \nu_n),(b,i \nu_n')} &= \tilde{G}^{\mathrm{PH1}}_{a b}(i\omega_n,i\nu_n,i\nu_n') - \delta_{i \omega_n,0} \tilde{G}_{a}(i\nu_n) \tilde{G}_{b}(i\nu_n')
\end{align}
with
\begin{equation}
\begin{aligned}
    &\tilde{G}^{\mathrm{PH1}}_{ab}(i\omega_n,i\nu_n,i\nu_n')= \\  \frac{1}{\beta^2} \iiiint_0^\beta d \tau_1 d \tau_2 d \tau_3 d \tau_4 e^{-i \nu_n(\tau_1-\tau_2)} &e^{-i \omega_n(\tau_2-\tau_3)} e^{-i \nu_n'(\tau_3-\tau_4)} \langle \mathcal{T}_\tau \hat{c}^{\dag}_{a}(\tau_1) \hat{c}_a(\tau_2) \hat{c}^{\dag}_{b}(\tau_3) \hat{c}_{b}(\tau_4) \rangle .
\end{aligned}
\end{equation}

We then perform the inversion
\begin{align}
    \underline{\chi}^{\mathrm{PH}}(Q) =  \Bigg[ \underline{\chi}^{(0,\mathrm{PH})-1}(Q) - \underline{\tilde{\chi}}^{(0,\mathrm{PH1})-1}(i \omega_n) + \underline{\tilde{\chi}}^{(\mathrm{PH1})-1}(i \omega_n) \Bigg]^{-1} .
\end{align}

The physical susceptibility for a Hermitian observable $O$ may then be computed as
\begin{align}
    \chi_{OO}(Q) = \frac{1}{\beta}  \sum_{i\nu_n,i\nu_n'} \sum_{ab} O_{aa} O_{bb} \chi_{aa,bb}^{\mathrm{PH}}(Q,i\nu_n,i\nu_n') , 
\end{align}
where $O_{aa} = O_{(A,\sigma)(A,\sigma)} = (-1)^{(1-\sigma)}$ for $zz$-spin correlations ($A$ and $\sigma=1,2$ are orbital and spin indices respectively). The orbital index can be ignored here, since we study a one-orbital model.

\subsection{Density and magnetic channel parameterization}

It can be shown that in the case of a $SU(2)$-symmetric system, only pairwise equal spin components of the susceptibilities are non-zero due to spin conservation. This in turn means that $\chi^{(\mathrm{PH})}$ can be written in two decoupled channels: $\chi^{(\mathrm{PH})}_{\sigma\sigma,\sigma'\sigma'} = \chi^{(\mathrm{PH})}_{\sigma\sigma'}$ (longitudinal spin channel) and $\chi^{(\mathrm{PH})}_{\sigma\sigma',\sigma'\sigma}=\chi^{(\mathrm{PH})}_{\sigma\bar{\sigma}'}$ (transverse spin channel). For the computation of physical observables, such as the $zz$-spin and charge susceptibilities, one needs only the longitudinal spin channel. 

The SU(2) symmetry furthermore implies $X_{\uparrow\uparrow}= X_{\downarrow\downarrow},X_{\uparrow\downarrow}= X_{\downarrow\uparrow}$, where $X=\chi,\chi^0,\Gamma$. We can use these relations to cast the BSE into a diagonal form for $X'=\mathrm{diag}(X^d,X^m)$, with $ X^d := X_{\uparrow\uparrow}+X_{\uparrow\downarrow}, X^m := X_{\uparrow\uparrow}-X_{\uparrow\downarrow}$, where the $d$ and $m$ superscripts refer to the density and magnetic channels, respectively. The BSE equations can then be solved separately in the two decoupled channels:
\begin{align}
    \tilde{\chi}^{0,d}(i \omega_n)_{(A,i \nu_n),(B,i \nu_n')} &= - \tilde{G}_{A \uparrow}(i\nu_n) \tilde{G}_{A \uparrow}(i\nu_n-i \omega_n) \delta_{i\nu_n,i\nu_n'} \delta_{AB} , \\
    \chi^{0,d}(\mathbf{q},i \omega_n)_{(A,i \nu_n),(B,i \nu_n')} &= - \frac{1}{N_k} \sum_{\mathbf{k}} G_{BA \uparrow}(\mathbf{k},i\nu_n) G_{AB \uparrow}(\mathbf{k}-\mathbf{q},i\nu_n - i \omega_n) \delta_{i\nu_n,i\nu_n'} ,
\end{align}
\begin{equation}
\begin{aligned}
    \tilde{\chi}^{d}(i \omega_n)_{(A,i \nu_n),(B,i \nu_n')} =& \,\tilde{G}^{\mathrm{PH1}}_{A \uparrow B \uparrow}(i\omega_n,i\nu_n,i\nu_n') + \tilde{G}^{\mathrm{PH1}}_{A \uparrow B \downarrow}(i\omega_n,i\nu_n,i\nu_n') \\ &- 2\delta_{i \omega_n,0} \tilde{G}_{A \uparrow}(i\nu_n) \tilde{G}_{B \uparrow}(i\nu_n') ,
\end{aligned}
\end{equation}
\begin{align}
    \tilde{\chi}^{0,m}(i \omega_n)_{(A,i \nu_n),(B,i \nu_n')} &= - \tilde{G}_{A \uparrow}(i\nu_n) \tilde{G}_{A \uparrow}(i\nu_n-i \omega_n) \delta_{i\nu_n,i\nu_n'} \delta_{AB} ,\\ 
    \chi^{0,m}(\mathbf{q},i \omega_n)_{(A,i \nu_n),(B,i \nu_n')} &= - \frac{1}{N_k} \sum_{\mathbf{k}} G_{BA \uparrow}(\mathbf{k},i\nu_n) G_{AB \uparrow}(\mathbf{k}-\mathbf{q},i\nu_n - i \omega_n) \delta_{i\nu_n,i\nu_n'} ,\\  
    \tilde{\chi}^{m}(i \omega_n)_{(A,i \nu_n),(B,i \nu_n')} &= \tilde{G}^{\mathrm{PH1}}_{A \uparrow B \uparrow}(i\omega_n,i\nu_n,i\nu_n') - \tilde{G}^{\mathrm{PH1}}_{A \uparrow B \downarrow}(i\omega_n,i\nu_n,i\nu_n'),
\end{align}
where the capital indices index the orbitals, and can  be ignored in the case of the one-orbital model. Note that the non-interacting susceptibilities are the same in both the density and magnetic channels.

\section{AFM phase}
\label{bse_afm_app}
We now consider a particular case with broken SU(2) symmetry, the AFM state. To stabilize an AFM solution of the Hubbard model on the square lattice requires the doubling of the unit cell, whose sites are indexed by the sublattice index $\lambda=1,2$. The DMFT self-consistency is modified in the following way, for the local impurity Green's function:
\begin{align}
    \tilde{G}_{\lambda=1,A\sigma}(i \omega_n) = \tilde{G}_{\lambda=2,A\bar{\sigma}}(i \omega_n).
\end{align}

For the local two-particle Green's function, we have
\begin{align}
    \tilde{G}^{\mathrm{PH}}_{\lambda=1,A\sigma;\lambda=1,B\sigma'}(i \omega_n,i\nu_n,i\nu_n') = \tilde{G}^{\mathrm{PH}}_{\lambda=2,A\bar{\sigma};\lambda=2,B\bar{\sigma}'}(i \omega_n,i\nu_n,i\nu_n'),  \label{G2_afm}
\end{align}
where the barred indices correspond to spin-flipped ones.

The lattice BSE equation expressed in a matrix form (in the sublattice indices) reads
\begin{align}
    \begin{pmatrix}
        \chi_{11} & \chi_{12} \\ \chi_{21} & \chi_{22}
    \end{pmatrix}_{\mathbf{q}} =  \begin{pmatrix}
        \chi^0_{11} & \chi^0_{12} \\ \chi^0_{21} & \chi^0_{22}
    \end{pmatrix}_{\mathbf{q}} + \frac{1}{\beta^2}\begin{pmatrix}
            \chi^0_{11} & \chi^0_{12} \\ \chi^0_{21} & \chi^0_{22}
    \end{pmatrix}_{\mathbf{q}} \begin{pmatrix}
        \tilde{\Gamma}_{11} & 0 \\ 0 & \tilde{\Gamma}_{22}
    \end{pmatrix} \begin{pmatrix}
            \chi_{11} & \chi_{12} \\ \chi_{21} & \chi_{22}
    \end{pmatrix}_{\mathbf{q}} , 
\end{align}
where each element is itself a matrix in the remaining orbital, spin and frequency indices, and we used the short-hand notation $\chi_{\lambda \lambda'} := \chi_{\lambda\lambda,\lambda'\lambda'}$. The corresponding local BSE becomes
\begin{align}
    \begin{pmatrix}
        \tilde{\chi}_{11} & 0 \\ 0 & \tilde{\chi}_{22}
    \end{pmatrix} =  \begin{pmatrix}
        \tilde{\chi}^0_{11} & 0 \\ 0 & \tilde{\chi}^0_{22}
    \end{pmatrix} + \frac{1}{\beta^2}\begin{pmatrix}
        \tilde{\chi}^0_{11} & 0 \\ 0 & \tilde{\chi}^0_{22}
    \end{pmatrix} \begin{pmatrix}
            \tilde{\Gamma}_{11} & 0 \\ 0 & \tilde{\Gamma}_{22}
    \end{pmatrix} \begin{pmatrix}
        \tilde{\chi}_{11} & 0 \\ 0 & \tilde{\chi}_{22}
    \end{pmatrix},
\end{align}
since we can only capture local terms. The BSE equations in the AFM phase read explicitly:
\begin{align}
    \tilde{\chi}^{0,\mathrm{AFM}}(i \omega_n)_{(\lambda,A,\sigma, i \nu_n),(\lambda,B,\sigma',i \nu_n')} &= - \tilde{G}_{\lambda A \sigma}(i\nu_n) \tilde{G}_{\lambda A \sigma}(i\nu_n-i \omega_n) \delta_{i\nu_n,i\nu_n'} \delta_{AB} \delta_{\sigma \sigma'},
\end{align}
\begin{equation}
\begin{aligned}
     \chi^{0,\mathrm{AFM}}(\mathbf{q},i \omega_n)_{(\lambda,A,\sigma, i \nu_n),(\lambda',B,\sigma',i \nu_n')} =& - \frac{1}{N_k} \sum_{\mathbf{k}} G_{\lambda' \lambda, BA, \sigma}(\mathbf{k},i\nu_n) G_{\lambda \lambda',AB, \sigma}(\mathbf{k}-\mathbf{q},i\nu_n - i \omega_n) \\ &\cdot \delta_{i\nu_n,i\nu_n'} \delta_{\sigma \sigma'},
\end{aligned}
\end{equation}
\begin{align}
     \tilde{\chi}^{\mathrm{AFM}}(i \omega_n)_{(\lambda,A,\sigma, i \nu_n),(\lambda,B,\sigma',i \nu_n')} &= \tilde{G}^{\mathrm{PH1}}_{\lambda A\sigma,\lambda B \sigma'}(i\omega_n,i\nu_n,i\nu_n') - \delta_{i \omega_n,0} \tilde{G}_{\lambda A \sigma}(i\nu_n) \tilde{G}_{\lambda B\sigma'}(i\nu_n'),
\end{align}
where the $\lambda$ indices index the sublattice sites. The physical susceptibility, for a Hermitian observable $O$, unfolded to the full one-site Brillouin zone of the square lattice becomes
\begin{align}
    \chi_{OO}(\mathbf{q}_{\mathrm{BZ}},i\omega_n) = \frac{1}{\beta}  \sum_{i\nu_n,i\nu_n'} \sum_{ab} e^{i \mathbf{q_{\mathrm{BZ}}}(\tau_{\lambda}-\tau_{\lambda'})} O_{aa} O_{bb} \chi_{aa,bb}^{\mathrm{PH}}(\mathbf{q}_{\mathrm{RBZ}},i\omega_n,i\nu_n,i\nu_n'),
\end{align}
where $O_{aa} = O_{(\lambda A \sigma),(\lambda A \sigma)} = (-1)^{(1-\sigma)}$ for $zz$-spin correlations ($\lambda=1,2$; $A$ and $\sigma=1,2$ are sublattice, orbital and spin indices, respectively). The orbital indices can be ignored for the 1-orbital model. The phase factor implements the unfolding to the full Brillouin zone and depends on the atomic positions in the unit cell, indexed by vectors $\tau_\lambda$.

\end{widetext}

\bibliographystyle{apsrev4-2}
\bibliography{hubbard_bse}

\end{document}